\documentclass[twocolumn,showpacs,preprintnumbers,amsmath,amssymb,superscriptaddress,prb]{revtex4}

\usepackage{graphicx}
\usepackage{dcolumn}
\usepackage{bm}

\begin{document}

\preprint{APS/123-QED}

\title{
Doping-Dependent and Orbital-Dependent Band Renormalization
in Ba(Fe$_{1-x}$Co$_x$)$_2$As$_2$ Superconductors
}

\author{T.~Sudayama}
\affiliation{Department of Physics, University of Tokyo, 5-1-5 Kashiwanoha, Kashiwa, 
Chiba 277-8561, Japan}
\author{Y.~Wakisaka}
\affiliation{Department of Physics, University of Tokyo, 5-1-5 Kashiwanoha, Kashiwa, 
Chiba 277-8561, Japan}
\author{T.~Mizokawa}
\affiliation{Department of Complexity Science 
and Engineering, University of Tokyo, 5-1-5 Kashiwanoha, Kashiwa, 
Chiba 277-8561, Japan}
\affiliation{Department of Physics, University of Tokyo, 5-1-5 Kashiwanoha, Kashiwa, 
Chiba 277-8561, Japan}
\author{S.~Ibuka}
\affiliation{Institute for Solid State Physics, University of Tokyo, 106-1 Shirakata, 
Tokai, Ibaraki 319-1106, Japan}
\affiliation{JST, TRIP, 5, Sanbancho, Chiyoda, Tokyo 102-0075, Japan}
\author{R.~Morinaga}
\affiliation{Institute for Solid State Physics, University of Tokyo, 106-1 Shirakata, 
Tokai, Ibaraki 319-1106, Japan}
\affiliation{JST, TRIP, 5, Sanbancho, Chiyoda, Tokyo 102-0075, Japan}
\author{T.~J.~Sato}
\affiliation{Institute for Solid State Physics, University of Tokyo, 106-1 Shirakata, 
Tokai, Ibaraki 319-1106, Japan}
\affiliation{JST, TRIP, 5, Sanbancho, Chiyoda, Tokyo 102-0075, Japan}
\author{M.~Arita}
\affiliation{Hiroshima Synchrotron Radiation Center, Hiroshima University, 
Higashihiroshima, Hiroshima 739-0046, Japan}
\author{H.~Namatame}
\affiliation{Hiroshima Synchrotron Radiation Center, Hiroshima University, 
Higashihiroshima, Hiroshima 739-0046, Japan}
\author{M.~Taniguchi}
\affiliation{Hiroshima Synchrotron Radiation Center, Hiroshima University, 
Higashihiroshima, Hiroshima 739-0046, Japan}
\affiliation{Graduate School of Science, Hiroshima University, 
Higashihiroshima, Hiroshima 739-8526, Japan}
\author{N.~L.~Saini}
\affiliation{Department of Physics, Universit\'a di Roma "La Sapienza",
Piazzale Aldo Moro 2, 00185 Roma, Italy}
\affiliation{Department of Complexity Science 
and Engineering, University of Tokyo, 5-1-5 Kashiwanoha, Kashiwa, 
Chiba 277-8561, Japan}

\date{\today}

\begin{abstract}
Angle resolved photoemission spectroscopy of
Ba(Fe$_{1-x}$Co$_x$)$_2$As$_2$ ($x$ = 0.06, 0.14, and 0.24) 
shows that the width of the Fe 3$d$ $yz$/$zx$ hole band 
depends on the doping level. In contrast, the Fe 3$d$ $x^2-y^2$ 
and $3z^2-r^2$ bands are rigid and shifted by the Co doping. 
The Fe 3$d$ $yz$/$zx$ hole band is flattened at the optimal
doping level $x$ = 0.06, indicating that the band renormalization
of the Fe 3$d$ $yz$/$zx$ band correlates with the enhancement 
of the superconducting transition temperature.
The orbital-dependent and doping-dependent band renormalization 
indicates that the fluctuations responsible for the superconductivity 
is deeply related to the Fe 3$d$ orbital degeneracy.
\end{abstract}

\keywords{Fe-based superconductors, angle-resolved photoemission 
spectroscopy, band renormalization, orbital degeneracy}
\maketitle

\newpage

\section{Introduction}

Research activities to understand fundamental mechanisms 
of high temperature superconductivity have been accelerated 
by the discovery of superconductivity in the FeAs systems
\cite{Kamihara2008, Takahashi2008}.
The FeAs-based superconductors commonly have the FeAs layers
where Fe atoms are tetrahedrally coordinated by As,
and are obtained by carrier doping to the antiferromagnetic 
parent compounds such as LaFeAsO and BaFe$_2$As$_2$.
BaFe$_2$As$_2$ shows superconductivity by electron doping 
with the highest $T_c$ of 25 K in Ba(Fe,Co)$_2$As$_2$ 
although the FeAs plane is highly disordered by the Co doping
\cite{Sefat2008, Nakajima2009}.
Since the superconducting phase is closely related to 
the magnetic phase in the FeAs-based as well as 
CuO-based high-$T_c$ superconductors, many experimental 
and theoretical studies have been dedicated to understand 
the relationship between the superconductivity and magnetism \cite{Lee}. 
When the 3$d$ electrons are localized to form a Mott insulator,
the symmetry of transition-metal 3$d$ orbitals controls 
the magnetic interaction between the transition-metal
3$d$ spins (Kugel-Khomskii mechanism), while the symmetry 
breaking of the transition-metal 3$d$ orbitals is accompanied 
by the local lattice distortion (Jahn-Teller mechanism) \cite{KK}.
When the 3$d$ electrons are itinerant, the band Jahn-Teller effect
coupled with lattice distortion can change the Fermi surface topology
to induce magnetic instability (Orbitally-induced Peierls mechanism)
\cite{KM}.
Actually, the anomalous lattice instability is commonly found 
in the various high-$T_c$ superconductors including 
the cuprate-based and Fe-based superconductors \cite{Saini2001,Joseph2010}. 
In this context, the transition-metal 3$d$ orbital degree of freedom 
is the key ingredient which bridges between the lattice instability 
and the magnetic instability both in localized and itinerant cases.

The angle-resolved photoemission spectroscopy (ARPES) is a powerful
technique to study the electronic structure of multi-orbital systems
where the lattice and magnetic instabilities may exist. 
As for the FeAs systems, ARPES studies on hole-doped
\cite{Ding2008, Liu2008, Zabolotnyy2009, Zhang2009}
and electron-doped 
\cite{Terashima2008, Sekiba2008, Malaeb2009, Vilmercati2009,
Sudayama2010, Thirupathaiah2010, Mansart2011}
BaFe$_2$As$_2$ have confirmed the importance 
of the multi-orbital character as predicted 
by the band-structure calculations 
\cite{Singh2008, Kuroki2008, Thirupathaiah2010}. 
However, the effect of Co doping on evolution of 
multi-orbital electronic structure is rather complicated 
\cite{Wadati2010} and should be examined further
to understand the fundamental mechanism of the superconductivity.
Here, we report an ARPES study on Ba(Fe$_{1-x}$Co$_x$)$_2$As$_2$
with $x$ = 0.06, 0.14, and 0.24 around $\Gamma$ point. 
It has been found that the upward chemical potential shift 
with the Co doping basically supports the electron doping picture. 
However, while the renormalization factor for the $3z^2-r^2$ 
and $x^2-y^2$ bands does not depend on the doping level,
that for the $yz$/$zx$ hole band decreases with the 
Co doping breaking the simple rigid band model. 
The band narrowing of the $yz$/$zx$ hole band correlates 
with the enhancement of $T_c$, indicating that the orbital
effect plays important roles for the mechanism of high-$T_c$ 
superconductivity.

Single crystals of Ba(Fe$_{1-x}$Co$_x$)$_2$As$_2$
with $x$ = 0.06, 0.14, and 0.24
were grown by Bridgman method with FeAs flux \cite{Morinaga2009}.
ARPES measurements were performed at beam line 9A, 
Hiroshima Synchrotron Radiation Center (HSRC) using a SCIENTA R4000 
analyzer with circularly polarized light ($h\nu$ = 17 eV and 23 eV).
Total energy resolutions were set to 18 meV and 14 meV
for $h\nu$ = 23 eV and 17 eV, respectively.
We cleaved the single crystals at 30 K under ultrahigh vacuum of
$5 \times10^{-9}$ Pa and the ARPES data of the cleaved surface 
parallel to the FeAs plane were collected at 30 K within four hours 
after the cleaving.

\begin{figure}
\includegraphics[width=0.45\textwidth]{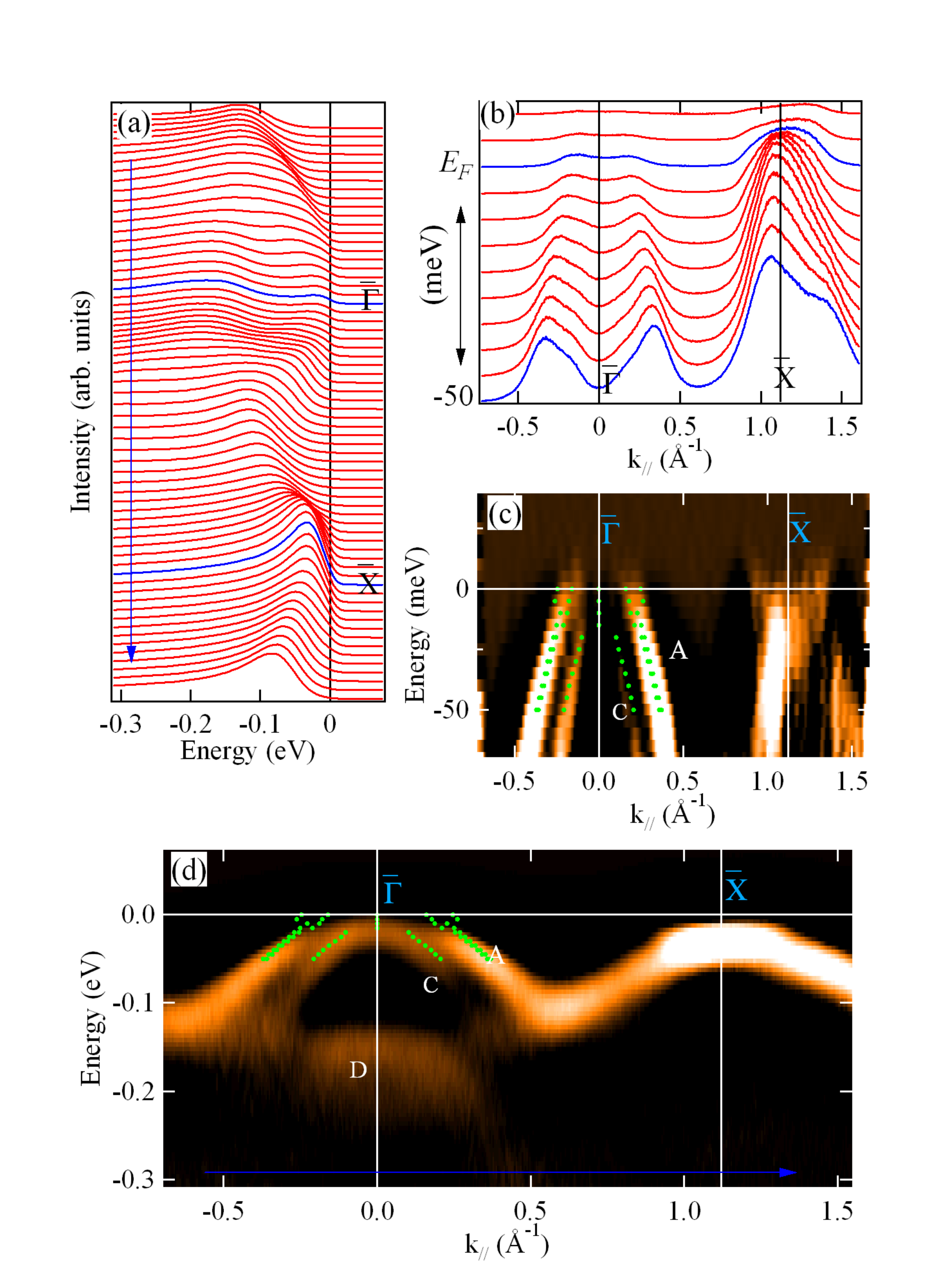}
\caption{(Color online) 
(a) EDC plot and (b) MDC plot 
for $x$=0.06 taken at $h\nu$ = 23 eV. 
(c) Second derivative plot of MDC. 
(d) Second derivative plot of EDC.
The dots indicate the band locations determined 
by fitting MDCs to Lorentzian functions.
The $\bar{\Gamma}$-$\bar{X}$ direction is the nearest neighbor 
Fe-Fe direction in the plane.
}
\end{figure}

\begin{figure}
\includegraphics[width=0.45\textwidth]{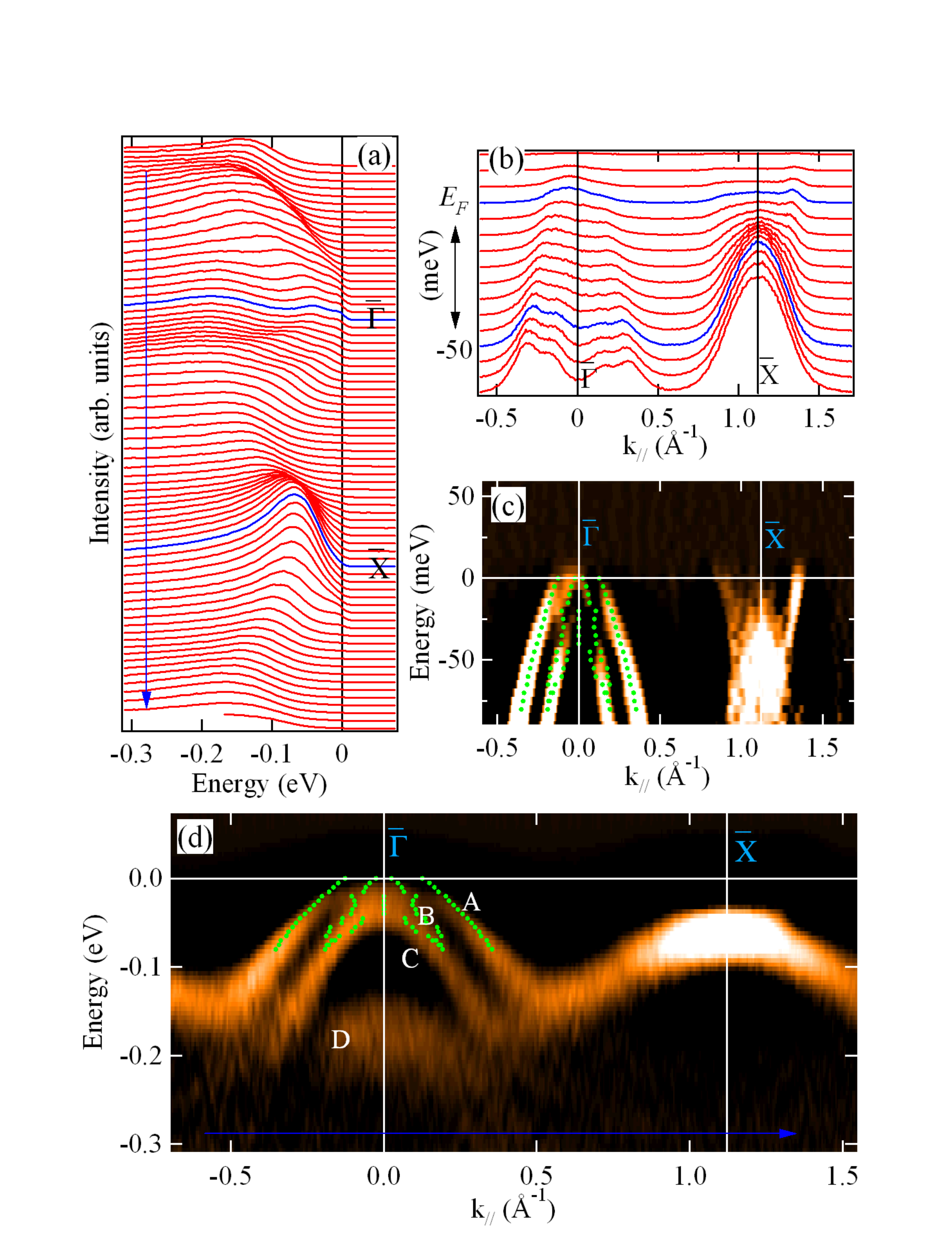}
\caption{(Color online) 
(a) EDC plot and (b) MDC plot 
for $x$=0.14 taken at $h\nu$ = 23 eV. 
(c) Second derivative plot of MDC. 
(d) Second derivative plot of EDC.
The dots indicate the band locations determined 
by fitting MDCs to Lorentzian functions.
The $\bar{\Gamma}$-$\bar{X}$ direction is the nearest neighbor 
Fe-Fe direction in the plane.
}
\end{figure}	

Figures 1(a) and (b) show energy distribution curve (EDC)
and momentum distribution curve (MDC) plots of the ARPES data 
for $x$=0.06 (optimally-doped system) taken at 
$h\nu$ = 23 eV from the zone center ($\bar{\Gamma}$ point) 
to the zone corner ($\bar{X}$ point) of the two-dimensional Brillouin zone.
The in-plane momentum $k_{\parallel}$ is swept along the nearest neighbor 
Fe-Fe direction in the plane.
At $h\nu$ = 23 eV, the $\bar{\Gamma}$ point corresponds to
the $\Gamma$ point of the three-dimensional 
Brillouin zone where the out-of plane momentum $k_{z}$ is zero. 
The second derivative plots of MDC and EDC are displayed 
in Figs. 1(c) and (d), respectively.
By comparing the present 
result with the band structure calculation near the $\Gamma$ point
\cite{Thirupathaiah2010}, the parabolic band crossing $E_F$ 
can be assigned to one of the $yz/zx$ bands (band A).
Here, the $z$ axis is perpendicular to the FeAs plane and 
the $x$ axis is along the nearest neighbor Fe-Fe direction 
in the plane.
On the other hand, the other hole band takes its maximum 
at $\sim$ -10 meV and can be assigned to the $x^2-y^2$ 
band (band C). The broad and flat band at $\sim$ -170 meV 
is identified as the $3z^2-r^2$ band (band D).
The observation of the two hole bands (bands A and C)
in the optimally-doped system is apparently inconsistent 
with that of the three hole bands in the overdoped system 
(bands A, B and C, see Fig. 2) \cite{Sudayama2010}. 
In order to further examine this apparent inconsistency, 
we have tried to determine the band locations of 
the two hole bands by fitting MDCs to Lorentzian functions. 
The results are shown by the dots in Figs. 1 (c) and (d).
Interestingly, the $yz/zx$ hole band (band A) near $E_F$ 
(in the region above -10 meV) is found to become broad 
in the momentum space compared to that below -10 meV.
Consequently, it is possible to fit the broad $yz/zx$ hole band 
to the two components in the energy region from $E_F$ to -10 meV
although the fitting to the two components is just
to demonstrate the broadening and is not a unique solution.
Below -10 meV, the two components merge to form 
the single yz/zx hole hand, indicating that band A 
becomes narrow below -10 meV compared to that above -10 meV.
The disappearance of band B can be attributed to the smearing 
effect of $k_z$ dispersion and/or the photoemission matrix element effect
\cite{Liu2008}. 
However, in the ARPES measurements under various conditions
\cite{ Terashima2008, Sekiba2008, Malaeb2009, Vilmercati2009,
Sudayama2010, Thirupathaiah2010, Mansart2011}, band B is always 
absent near the $\Gamma$ point, suggesting that the absence 
of band B is not due to the smearing effect of $k_z$ dispersion 
and/or the matrix element effect.

EDC and MDC plots for $x$ = 0.14 (overdoped system) taken 
at $h\nu$ = 23 eV are displayed in Figs. 2(a) and (b), respectively. 
Figures 2(c) and (d) shows the second derivative plots 
made up from the MDC and EDC data, respectively.
Three parabolic bands can be identified around the zone center
and two of them (bands A and B) reach $E_F$. In this plot, 
the three hole bands are identified by fitting MDCs
to Lorentzian functions as already reported in the literature
\cite{Sudayama2010}.
By comparing the EDC data in Fig. 2(a) to that in Fig. 1(a),
the hole bands at the zone center are sharpened and
the electron bands at the zone corner are broadened in going 
from the optimally-doped system to the overdoped system.

\begin{figure}
\includegraphics[width=0.45\textwidth]{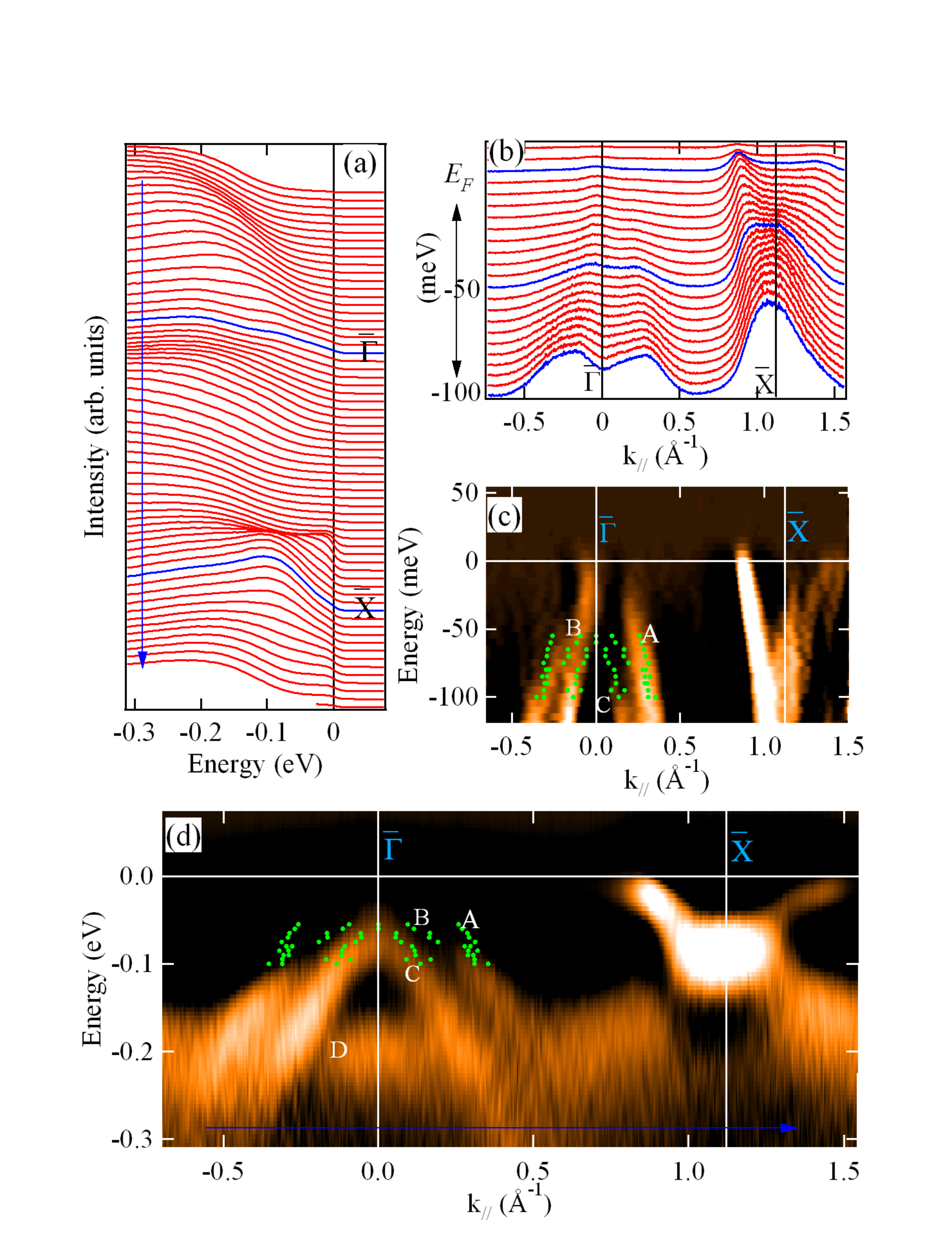}
\caption{(Color online)
(a) EDC plot and (b) MDC plot 
for $x$=0.24 taken at $h\nu$ = 23 eV. 
(c) Second derivative plot of MDC. 
(d) Second derivative plot of EDC.
The dots indicate the band locations determined 
by fitting MDCs to Lorentzian functions.
The $\bar{\Gamma}$-$\bar{X}$ direction is the nearest neighbor 
Fe-Fe direction in the plane.
}
\end{figure}

\begin{figure}
\includegraphics[width=0.45\textwidth]{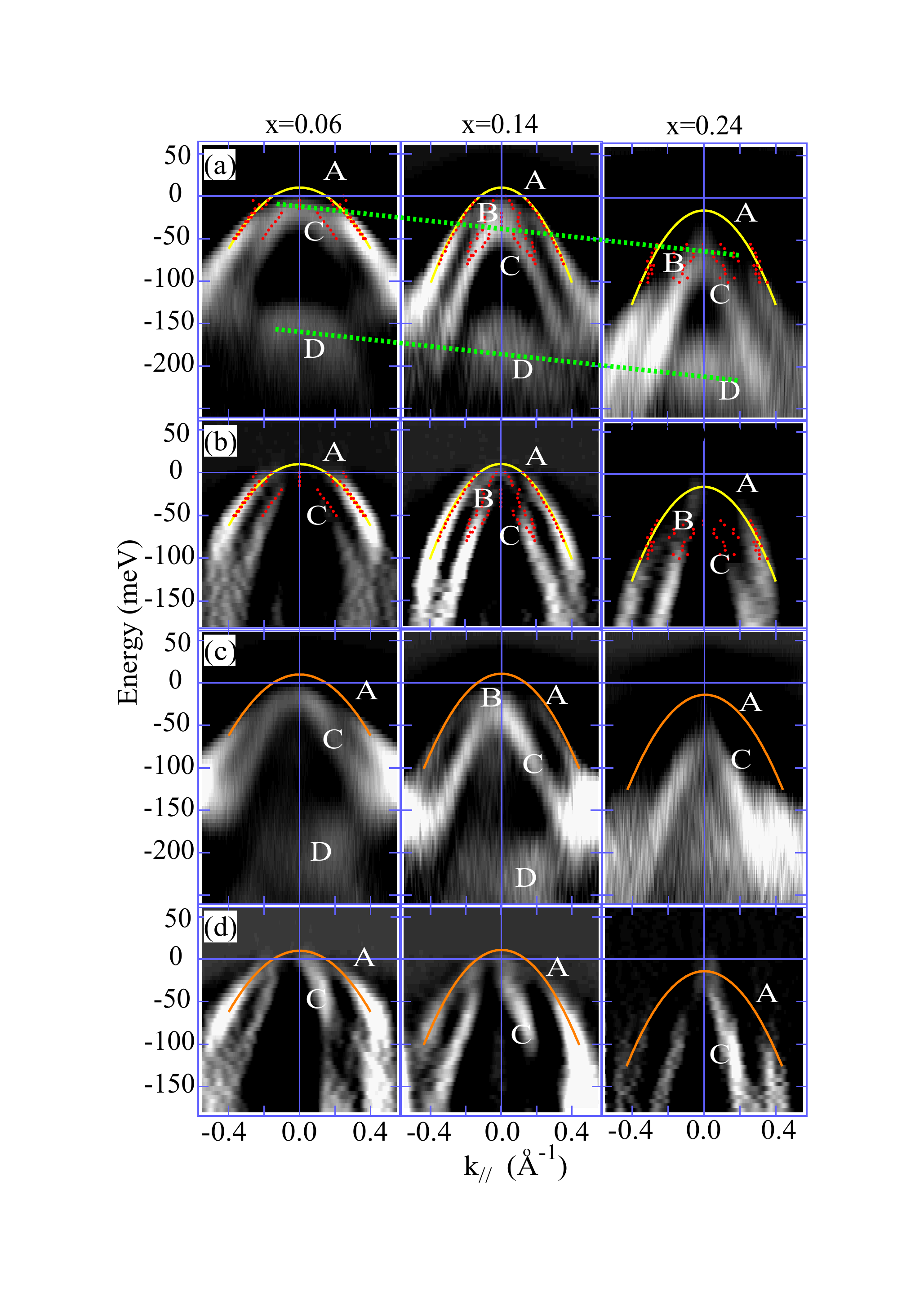}
\caption{(Color online)
(a) Second derivative plots for $x$=0.06, 0.14, and 0.24 
at $h\nu$ = 23 eV.
The dots indicate the band locations 
determined by fitting the MDCs to Lorentzian functions. 
The solid parabolic curves roughly 
show the dispersion of hole band A. The dotted lines
roughly show the energy shift of top of hole band C
and that of band D.
(b) Second derivative plots of MDC for $x$=0.06, 0.14, and 0.24 at $h\nu$ = 23 eV.
(c) Second derivative plots of EDC for $x$=0.06, 0.14, and 0.24 at $h\nu$ = 17 eV.
(d) Second derivative plots of MDC for $x$=0.06, 0.14, and 0.24 at $h\nu$ = 17 eV.
}
\end{figure}

Figures 3(a) and (b) show EDC and MDC plots for $x$ = 0.24 
(heavily-overdoped system) taken at $h\nu$ = 23 eV, respectively.
In the heavily-overdoped system, the hole bands at the zone center 
are very broad in the EDC plot probably because the hole bands are
fully occupied by electrons and are located well below $E_F$.
Assuming that three hole bands should exist, the broad MDC data 
can be fitted to the model Lorentzian functions with three components.
In the second derivative plots displayed in Figs. 3(c) and (d), 
the band locations of the three components determined by the fitting 
are plotted. In the heavily-overdoped case, the hole bands never 
reach $E_F$, consistent with the electron doping from $x$=0.14 to $x$=0.24
and the result of neutron scattering experiment \cite{Matan2010}.

The second derivative plot of MDC is useful to identify 
the dispersive bands while the dispersionless band 
can be found in the second derivative plot of EDC.
In Figs. 4 (a) and (b), the second derivative plots 
of EDC and MCD for the hole bands around the $\Gamma$ point
are displayed in order to show the evolution 
of the hole band dispersions as a function of the Co doping. 
Firstly, the top of hole band C (the $x^2-y^2$ band)
is located at -10 meV, -30 meV, and -50 meV
for $x$ = 0.06, 0.14, and 0.24, respectively.
This energy shift [indicated by the dotted line in Fig. 4(a)]
is consistent with the upward chemical potential shift 
by the electron doping. The flat band at $\sim$ -170 meV (band D)
can be assigned to the renormalized $3z^2-r^2$ band 
which is located at $\sim$ -700 meV in the LDA calculation
\cite{Singh2008}. Band D also shows the same energy shift
as band C. The discrepancy between the experimental results
and the band-structure calculations can be explained by
the band renormalization due to correlation effects.
Figure 4 shows that the degree of the band renormalizations 
for the $x^2-y^2$ and $3z^2-r^2$ bands does not depend on 
the doping level. 
These two bands show rigid band energy shift 
due to the chemical potential shift by the Co doping.
Secondly, the slope of hole band A (one of the $yz/zx$ bands) 
increases from the optimally doped system to the overdoped system.
In Fig. 4, the dispersion of hole band A is roughly shown
by the solid parabolic curves which are given by $E = ak^2+b$.
Here, $E$ and $k$ are energy (in unit of eV) and momentum 
(in unit of $\AA$) of the electron, respectively.
In the optimally doped system, $a = -0.45$ and $b = 0.01$,
while $a = -0.70$ ($-0.70$) and $b = 0.01$ ($-0.015$) for 
the overdoped (heavily-overdoped) system.
In the region of $0.25 \AA^{-1} < k_{//} < 0.35 \AA^{-1}$,
the energy difference between the highest and the lowest energy
states are about 25 meV and 40 meV for x=0.06 and x=0.14, respectively.
Considering the accuracy of energy (less than 1 meV) and  $k_{//}$ 
(less than 0.01 $\AA^{-1}$), one can safely conclude that the width 
of hole band A is considerably reduced in the optimally doped system.
Figures 4 (c) and (d) show the second derivative plots 
of EDC and MCD taken at $h\nu$ = 17 eV for the hole bands 
at the zone center. 
At $h\nu$ = 17 eV, the $\bar{\Gamma}$ point corresponds to
the midpoint of $\Gamma$ and Z points of the three-dimensional 
Brillouin zone. The hole bands observed at $h\nu$ = 17 eV
are slightly flattened compared to those at $h\nu$ = 23 eV
due to the difference of the out-of plane momentum $k_{z}$.
The dispersions of hole band A are roughly given by 
$a = -0.45$ and $b = 0.01$, $a = -0.58$ and $b = 0.01$,
and $a = -0.58$ and $b = -0.015$ for the optimally doped, 
overdoped doped, and heavily-overdoped systems.
Although the width of hole band A decreases from $\Gamma$ to Z
in the overdoped system, it is further reduced in going from 
the overdoped system to the optimally doped system.

The above ARPES results show that the $yz$/$zx$ hole bands 
are strongly modified at the optimal doping level.
The band deformation would be related to the recent theoretical 
and experimental works on the essential role of orbital degrees 
of freedom in the Fe-based superconductors
\cite{Kontani2010,Lee2010}.
The tetragonal-to-orthorhombic structural transition 
temperature is extrapolated to 0 K at the optimal doping.
Therefore, one can speculate that the doping dependence 
of the orbital-selective band renormalization
is related to the orthorhombic lattice fluctuation,
which can couple with the $yz$/$zx$ orbitals and 
can provide inhomogeneity of orbital and lattice. 
Further theoretical and experimental studies are required to 
understand whether the orbital or lattice fluctuation really
contributes to the pairing mechanism or not.
However, at least, this type of orbital or lattice fluctuation 
is expected to enhance the spin fluctuation via the 
orbitally-induced Peierls coupling \cite{KM} or the orbitally-induced 
excitonic coupling \cite{Mizokawa2008}.

In conclusion, the angle resolved photoemission spectroscopy study
on Ba(Fe$_{1-x}$Co$_x$)$_2$As$_2$ ($x$ = 0.06, 0.14, and 0.24)
reveals the orbital- and doping-dependent band renormalization.
In the optimally-doped system,
the renormalization factor of the observed $yz$/$zx$ hole band is 
strongly enhanced compared to that in the overdoped regimes.
In contrast, the $x^2-y^2$ and $3z^2-r^2$ bands show 
rigid band energy shift due to the Co doping and their 
renormalization factors do not depend on the doping level. 
The orbital- and doping-dependent band renormalization indicates
that the orbital degree of freedom of the $yz$/$zx$ hole bands 
may play an essential role to enhance superconducting transition 
temperature.

This work was supported in part by the Global COE Program 
"the Physical Sciences Frontier", MEXT, Japan. 
The synchrotron radiation experiments have been done 
with the approval of HSRC (Proposals No. 09-A-25 and No. 10-A-10)

\end{document}